\documentclass[reprint,superscriptaddress,amsmath,amssymb,aps,pra,nofootinbib]{revtex4-2}

\usepackage{soul}
\usepackage{textcomp}  
\usepackage{graphicx}  
\usepackage{hyperref}  
\usepackage{siunitx}
\usepackage{xspace}
\usepackage{tabularx}
\usepackage[symbol]{footmisc}
\usepackage{IEEEtrantools}

\DeclareSIUnit\gauss{G}
\DeclareSIUnit\rad{rad}

\newcommand{\figref}[2][]{Fig.\,\ref{#2}#1}

\newcommand{\PRL}{Phys. Rev. Lett.}
\newcommand{\NaturePhysics}{Nat. Phys.}
\newcommand{\PRX}{Phys. Rev. X}
\newcommand{\PRA}{Phys. Rev. A}

\newcommand{\NatComm}{Nat. Commun.} 

\newcommand{\PRXQ}{PRX Quantum}
\newcommand{\JournalofPhysicsB}{J. Phys. B}
\newcommand{\AppliedPhysicsB}{Appl. Phys. B}

\makeatletter
\def\maketitle{
	\@author@finish
	\title@column\titleblock@produce
	\suppressfloats[t]}
\makeatother

\begin{document}
	
    \newcommand{\partitle}[1]{\section{#1}}

    \newcommand{\papertitle}{Dynamical spatial light modulation in the ultraviolet spectral range}

	\title{\papertitle{}}

    \author{Maximilian Ammenwerth}\thanks{These authors contributed equally to this work.}
	   \affiliation{Max-Planck-Institut f\"{u}r Quantenoptik, 85748 Garching, Germany}
	   \affiliation{Munich Center for Quantum Science and Technology (MCQST), 80799 Munich, Germany}

    \author{Hendrik Timme}\thanks{These authors contributed equally to this work.}
	   \affiliation{Max-Planck-Institut f\"{u}r Quantenoptik, 85748 Garching, Germany}
	   \affiliation{Munich Center for Quantum Science and Technology (MCQST), 80799 Munich, Germany}

    \author{Veronica Giardini}
	   \affiliation{European Laboratory for Non-Linear Spectroscopy (LENS), University of Florence, Via N. Carrara 1, 50019 Sesto Fiorentino, Italy}
       \affiliation{Department of Physics and Astronomy, University of Florence, Via G. Sansone 1, 50019 Sesto Fiorentino, Italy}

    \author{Renhao Tao}
	   \affiliation{Max-Planck-Institut f\"{u}r Quantenoptik, 85748 Garching, Germany}
	   \affiliation{Munich Center for Quantum Science and Technology (MCQST), 80799 Munich, Germany}
	   \affiliation{Fakultät für Physik, Ludwig-Maximilians-Universit\"{a}t, 80799 Munich, Germany}

    \author{Flavien Gyger}
	   \affiliation{Max-Planck-Institut f\"{u}r Quantenoptik, 85748 Garching, Germany}
	   \affiliation{Munich Center for Quantum Science and Technology (MCQST), 80799 Munich, Germany}

    \author{Ohad Lib}
	   \affiliation{Max-Planck-Institut f\"{u}r Quantenoptik, 85748 Garching, Germany}
	   \affiliation{Munich Center for Quantum Science and Technology (MCQST), 80799 Munich, Germany}

    \author{Dirk Berndt}
      \affiliation{Fraunhofer Institute for Photonic Microsystems IPMS, 01109 Dresden, Germany}

    \author{Dimitrios Kourkoulos}
        \affiliation{Fraunhofer Institute for Photonic Microsystems IPMS, 01109 Dresden, Germany}

    \author{Tim Rom}
        \affiliation{Fraunhofer Group for Microelectronics, Research Fab Microelectronics Germany (FMD), 10178 Berlin, Germany}

    \author{Immanuel Bloch}
	   \affiliation{Max-Planck-Institut f\"{u}r Quantenoptik, 85748 Garching, Germany}
	   \affiliation{Munich Center for Quantum Science and Technology (MCQST), 80799 Munich, Germany}
	   \affiliation{Fakultät für Physik, Ludwig-Maximilians-Universit\"{a}t, 80799 Munich, Germany}

    \author{Johannes Zeiher}
    \email{johannes.zeiher@mpq.mpg.de}
     \affiliation{Max-Planck-Institut f\"{u}r Quantenoptik, 85748 Garching, Germany}
     \affiliation{Munich Center for Quantum Science and Technology (MCQST), 80799 Munich, Germany}
     \affiliation{Fakultät für Physik, Ludwig-Maximilians-Universit\"{a}t, 80799 Munich, Germany}
	
	\date{\today}
	
	\begin{abstract}
        Spatial light modulators enable arbitrary control of the intensity of optical light fields and facilitate a variety of applications in biology, astronomy and atomic, molecular and optical physics. 
        For coherent light fields, holography, implemented through arbitrary phase modulation, represents a highly power-efficient technique to shape the intensity of light patterns. 
        Here, we introduce and benchmark a novel spatial light modulator based on a piston micro-mirror array.
        In particular, we utilize the reflection-based device to demonstrate arbitrary beam shaping in the ultraviolet regime at a wavelength of \SI{322}{\nano\meter}. 
        We correct aberrations of the reflected wavefront and show that the modulator does not add detectable excess phase noise to the reflected light field.
        We utilize the intrinsically low latency of the architecture to demonstrate fast switching of arbitrary light patterns synchronized with short laser pulses at an update rate of \SI{1}{\kilo\hertz}.
        Finally, we outline how the modulator can act as an important component of a zone-based architecture for a neutral-atom quantum computer or simulator, including ultraviolet wavelengths. 
    \end{abstract}
    \maketitle
	\section{Introduction}
    Spatial and dynamical light-field control is a ubiquitous experimental capability in atomic, molecular and optical physics.
    As one notable example, tweezer arrays of tightly focused laser beams have emerged as a powerful platform for quantum science experiments with neutral atoms, trapped ions and molecules~\cite{Saffman2016,Browaeys2020,Kaufman2021,Morgado2021,Schlosser2001,Barredo2016,Endres2016, Cooper2018, Norcia2018a, Saskin2019,Young2020a,Ebadi2021,Bluvstein2022,Ma2023,Evered2023a,Pause2024,Manetsch2024,Guttridge2024}.
    Various approaches have been followed for the generation of optical tweezers, including dynamical acousto-optic deflectors (AOD), static microlense arrays~\cite{Schffner2024}, meta-materials~\cite{Hsu2022,Holman2024,Huft2022}, digital micromirror devices (DMD)~\cite{Wang2020} or phase-only spatial light modulators.
    The latter is most widely used due to its appealing features such as power efficiency, high optical quality and control, combined with the possibility to dynamically adjust the optical potential during an experimental sequence~\cite{Lin2024, Knottnerus2025}.
    So far, the vast majority of experiments utilize visible light for generating optical traps. 
    This limitation is inherent to liquid-crystal-on-silicon spatial light modulators (LCOS-SLMs), the most widely-used technology available for phase-only spatial light modulation.
    \begin{figure}[b!]
	\includegraphics[width = 0.45\textwidth]{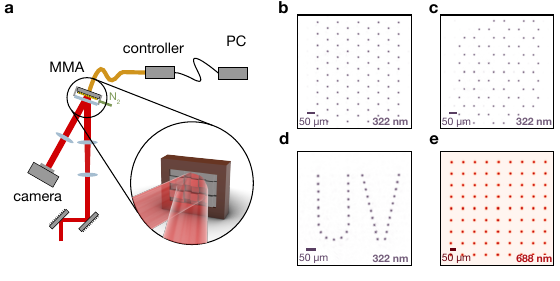}
        \caption{
            \textbf{Beam shaping with a piston micro-mirror array.} \textbf{a} The optical setup to generate configurable beam shapes with a phase-only spatial light modulator based on a micro-mirror array (MMA) is shown. The individual mirrors are axially displaced by up to \SI{350}{\nano\meter}%
            , thereby realizing an adjustable phase delay which is imprinted onto a laser beam reflected from the device. We characterize the generated patterns on a camera.
            \textbf{b-d} We benchmark the device at a wavelength of \SI{322}{\nano\meter} and demonstrate various tweezer arrays: triangular (b), honeycomb (c), and in a UV-letter shape (d) and reach state-of-the-art normalized homogeneity across the tweezer arrays of \SI{0.3}{\percent} peak-to-peak variation of the intensity.
            \textbf{e} Tweezer array at a wavelength of \SI{688}{\nano\meter} with 9x9 sites also reaching normalized peak-to-peak homogeneity of the intensity of \SI{0.3}{\percent}.
        }
		\label{fig:1}
	\end{figure}
    However, this technology is incompatible with modulation of light in the UV spectral range below approximately \SI{350}{\nano\meter} due to liquid crystal degradation under UV light exposure~\cite{Wen2005}.
    An alternative approach is based on micro-electromechanical system (MEMS) phase-only modulators, where the optical phase is controlled through the position of each micro-mirror in an array.
    %
    Recently, advanced spatial light-shaping capabilities and dynamical control of this approach have been demonstrated in the visible spectral range~\cite{Rocha2024,Rocha2025}.
    Compared with conventional approaches based on LCOS-SLMs, this technology is fully compatible with operation at UV wavelengths~\cite{Hacker2003}.
    This opens interesting perspectives for arbitrary dynamical spatial light-control in the UV, which is highly relevant for new applications in trapping atoms at ultra-small distances~\cite{Topcu2016}, or for homogenizing UV beams for zone-based quantum computing and quantum simulation architectures~\cite{Bluvstein2023}. 
    
    Here, we follow this route and extend the working range of phase-only spatial light modulation into the ultraviolet spectral range using a reflection-based micro-mirror array (MMA)~\cite{Rocha2024,Rocha2025}.
    We characterize the performance of the MMA at a wavelength of \SI{688}{\nano\meter}, demonstrating arbitrary optical potential shaping, low phase-noise operation and high optical power efficiency of the device.
    Furthermore, we implement arbitrary beam shape control in the ultraviolet regime at \SI{322}{\nano\meter}.
    Using spatially-resolved interferometric optical path length measurements, we characterize the wavefront of the reflected light and reach a root mean square (RMS) flatness of $\lambda/100$ at a wavelength of \SI{322}{\nano\meter}.
    Starting from this aberration-corrected beam we realize several different tweezer array configurations with a residual intensity non-uniformity smaller than \SI{0.3}{\percent} after equalizing the amplitudes of the tweezers using camera feedback ~\cite{Kim2019}.
    We shape the incoming light into flat-top beam profiles with excellent intensity homogeneity of as little as $0.75\%$ root mean square and $3.65\%$ peak-to-peak variation over a distance of \SI{113}{\micro\meter}.
    To demonstrate dynamical control, we synchronize the deflection of the MMA with short laser pulses and show configurable patterns at an update rate of \SI{1}{\kilo\hertz}, currently only limited by the electronic circuitry of the device.
    %
    %
    \begin{figure}[t]
		\centering
		\includegraphics{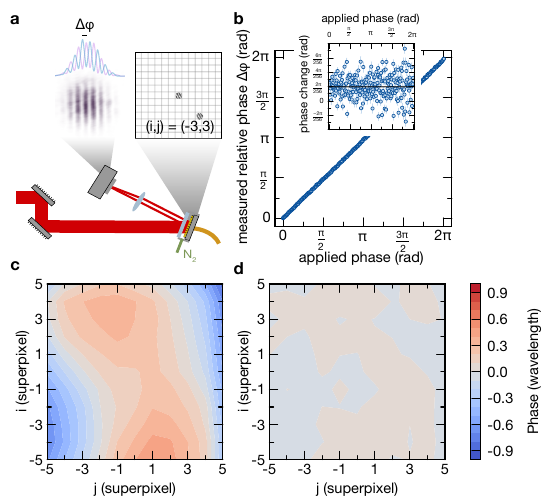}
		\caption{
			\textbf{Phase shifting interferometry.} 
			\textbf{a} To correct optical aberrations in the setup, we carry out phase-shifting interferometry by sequentially activating various superpixels on the MMA display. By varying the phase of one superpixel with respect to a common (and fixed) reference superpixel in the center of the MMA, we map out the phase mask required for aberration correction. We measure the interference fringe on the camera and fit the collapsed one-dimensional fringe to extract the phase.
            \textbf{b} We choose on reference superpixel (-3,3) and step the phase of the grating from $[0, 2\pi)$ in steps of $\frac{2\pi}{256}$. 
            We observe a linear relationship of the measured phase with respect to the applied phase, confirming the average phase step of $\frac{2\pi}{256}$ (inset).
            \textbf{c} The unwrapped and interpolated phase with the gradient removed is shown as a function of the superpixel index.
            \textbf{d} After the first round of wavefront measurement, we apply the measured phase as correction and repeat the wavefront error measurement using the same technique. The result is a flat phase profile verifying that the correction is effective.
        }
		\label{fig:2}
    \end{figure}
    \section{Beam shaping with a micro-mirror array}
    For our experiments, we use a prototype of a micro-mirror array system (\textit{64k customer evaluation kit piston MMA $\alpha$-module}) with $256 \times 256$ square mirrors with a pitch of \SI{16}{\micro\meter}.
    Each mirror is individually controlled electrostatically to configure its piston stroke with a resolution of 8-bits and a maximal stroke of \SI{350}{\nano\meter}.
    Starting with a flat phase profile of the incoming beam, a programmable phase pattern is imprinted onto the reflected beam by tailoring the stroke of the mirror array.
    The MMA is illuminated at an incident angle of approximately \SI{7}{\degree} and reflects the beam, see~\figref{fig:1}{a}.
    %
    In this configuration, the imprinted phase profile corresponds to twice the path length difference given by the mirror stroke, such that a full phase modulation contrast of $\phi(x,y) \in [0,2\pi)$ is feasible for wavelengths up to \SI{700}{\nano\meter}.
    We continuously flush the MMA surface with a weak flow of nitrogen to prevent damage to the micromechanical components of the chip from ozone, which forms from ambient oxygen under intense UV irradiation.
    %
    \subsection{Phase shifting interferometry}
    Generating diffraction-limited potentials requires the compensation of optical aberrations in the system.
    We employ phase-shifting interferometry to obtain a spatially resolved stationary phase correction pattern, which reflects the optical aberrations in the entire optical setup and can be used to correct the system ~\cite{Zupancic2016,Tsevas2021}.
    To obtain the phase correction map, we display a blazed grating on two circular patches of the MMA, one acting as reference, the other one as probe superpixel, each with a radius of 10 pixels.
    Both superpixels diffract the light of the incoming beam along different paths through the optical system, which results in an interference fringe on the camera; see~\figref{fig:2}{a}.
    The relative phase picked up by the beamlets emanating from the two superpixels translates to the phase of an interference fringe on the camera, which can be extracted directly.
    Keeping the position and phase of the reference pixel located in the center of the MMA fixed and varying the second superpixel's phase and position, we map out the entire spatially dependent phase profile due to aberrations in the system.
    In our setup, the initial wavefront error is dominated by astigmatism; see~\figref{fig:2}{c}.
    Subsequently, we correct the phase front by adding the corresponding correction pattern and characterize the phase front with another round of phase shifting interferometry.
    After one round of phase correction, we reach a state-of-the-art RMS flatness of $\lambda/100$ at our wavelength of \SI{322}{\nano\meter} across a circular aperture with the residual curvature shown in~\figref{fig:2}{d}.
    To create arbitrary beam shapes such as the structures shown in~\figref{fig:1}{b-e} we compute a hologram for the desired intensity pattern and measure the resulting intensity homogeneity on the camera~\cite{Chew2024}.
    Finally, we perform another round of optimizations of the hologram to equalize the tweezer intensity by weighing the amplitudes of the individual tweezers based on the measured image. 
    The intensity patterns created in this way reach an inhomogeneity (standard deviation over mean) of \SI{0.3}{\percent} both in the visible and UV spectral range; see~\figref{fig:1}{b-e}. 
    To characterize the phase resolution of the MMA, we step the phase of one superpixel in steps of $\frac{2\pi}{256}$ from $[0, 2\pi)$. 
    The interferometrically measured phase confirms the average step size of $\frac{2\pi}{256}$, see~\figref{fig:2}{b}.
    %
    %
    %
    The fluctuations of the measured phases indicate that the resolution of the optically realized hologram is reduced compared to the electronic resolution.
    %
    %
    From the deviation to a linear fit we calculate the integral nonlinearity as the largest absolute value of the residuals as $2.2 \times \frac{2 \pi}{256}$ ~\cite{Sansen2006} and obtain a standard deviation of $0.91 \times \frac{2 \pi}{256}$.
    %
    \section{MMA phase noise and timing}
    \begin{figure}[t!]
		\centering
		\includegraphics{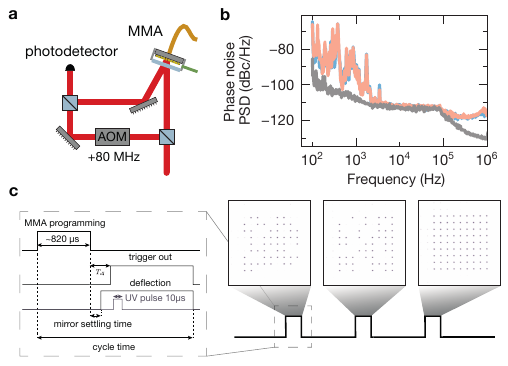}
        \caption{
			\textbf{Phase noise characterization and MMA timing} 
			\textbf{a} To measure the phase noise imprinted by the MMA, we set up an interferometer and frequency shift the reference arm by 80 MHz using an acousto-optic modulator (AOM). We interfere the frequency-shifted light with the light reflected from the MMA and detect the beat signal on a photodetector. 
            \textbf{b} For the phase noise measurement we program a constant displacement on the MMA. The phase noise is recorded with an electrical spectrum analyzer (Keysight N9020B), which is synchronized with the MMA. We observe no noticeable difference of the phase noise with activated MMA deflection (blue) compared to deactivated deflection (orange). The phase noise of the radio-frequency signal used to drive the AOM is shown in grey.
            \textbf{c} The typical timing cycle of the MMA is shown starting with the programming duration of \SI{820}{\micro\second}. 
            After the mirror settling time of approximately \SI{30}{\micro\second}, a user-configurable output trigger indicates that the desired deflection pattern is reached. 
		}
		\label{fig:3}
    \end{figure}
    \begin{figure*}[t!]
		\centering
		\includegraphics[width=\textwidth]{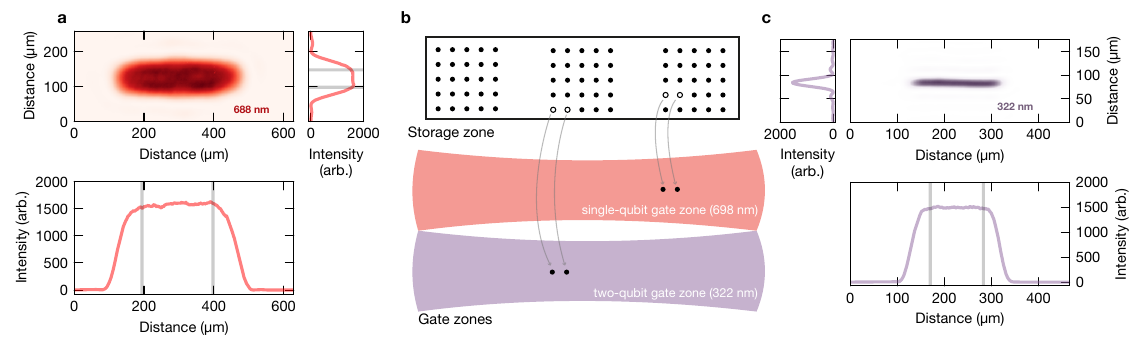}
        \caption{
            \textbf{Flat-top beam generation}
            \textbf{a} Generated flat-top beam profile at a wavelength of \SI{688}{\nano\meter}.
            \textbf{b} Overview of the zone architecture for neutral-atom-based quantum information processors. Arrays of neutral atoms are dynamically shuttled between dedicated zones for storage and for driving single- and two-qubit gates where a homogeneous flat-top illumination is highly advantageous. Additionally, individual atoms are addressed through a high-resolution objective. Executing programmable quantum circuits require rapid switching of the addressing pattern.
            \textbf{c} Generated flat-top beam profile at a wavelength of \SI{322}{\nano\meter}.
            }
		\label{fig:4}
    \end{figure*}
    \subsection{Phase noise of the MMA}
    Applications of arbitrary light patterns for trapping and manipulating individual atoms are particularly sensitive to noise in optical potentials~\cite{Savard1997}.
    As each mirror of the MMA is mounted on a spring-like structure and is controlled electrostatically to set the piston stroke, the noise characteristics need to be carefully analyzed to assess the range of possible applications of MMAs.
    To benchmark the stability of the displayed phase pattern, we measure the phase noise power spectral density with a Mach-Zehnder interferometer; see~\figref{fig:3}{a}.
    A collimated input beam at a wavelength of \SI{688}{\nano\meter} is split into two paths using a beam splitter.
    One arm of the interferometer impinges on the MMA, which displays a flat phase profile with a stroke of $\pi$.
    A second beam splitter is used to recombine the light with the reference path to superimpose both light fields on a photodetector.
    To avoid masking the signal in low-frequency noise, we frequency-offset the reference path by \SI{80}{\mega\hertz} using an acousto-optic modulator (AOM) and derive the relevant signal from the resulting AC signal on the photodetector.
    Phase noise of the MMA modulates the phase of the beat signal, which we characterize using an electrical phase noise analyzer.
    We configure a deflection duration of \SI{500}{\milli\second} and trigger the sweep of the phase noise analyzer during activated MMA deflection.
    As a reference, we repeat the measurement with the MMA in the off-state.
    Comparing the two measurements directly, we observe no noticeable excess noise when the MMA deflection is active; see~\figref{fig:3}{b}.
    Our measurement thus yields an upper bound to the phase noise of the MMA, which is negligible compared to the phase noise induced by mechanical instabilities of the optomechanical setup up to frequencies of a few~\si{\kilo\hertz}.
    At higher frequencies in the \SI{10}{\kilo\hertz} to \SI{100}{\kilo\hertz} regime, our measurement is limited by phase noise on the radio-frequency signal used for driving the AOM, but we do not expect significant noise contributions from the MMA in this frequency range.
    We conclude that the phase noise of the MMA is compatible with typical optical setups used for holographic tweezer generation and, thus, will not limit the achievable performance of the device.
    In particular, the characteristic switching noise of LCOS-SLMs is not present in this architecture, which might have beneficial consequences for extremely sensitive applications.
    \subsection{Timing and duty cycle of the MMA}
    Synchronization of the MMA deflection is possible via TTL inputs and outputs of the MMA controller.
    The MMA deflection cycle starts with a \SI{820}{\micro\second} long programming duration where the stroke of each mirror is configured.
    After programming, the device indicates the \textit{ready} status with a corresponding TTL output, and the deflection becomes active after a mirror settling time of approximately \SI{30}{\micro\second}.
    %
    %
    Synchronization with additional devices is possible via a user-configurable TTL output signal indicating the completed deflection of the MMA.
    To demonstrate rapid switching of phase patterns, we perform a proof-of-concept experiment in which we synchronize short laser pulses with the deflection of the MMA.
    At an update rate of \SI{1}{\kilo\hertz}, we apply light pulses of \SI{10}{\micro\second} duration controlled via an acousto-optic deflector.
    For these settings, corresponding to a duty cycle of \SI{1}{\percent}, we observe no degradation of the MMA for peak powers up to \SI{100}{\milli\watt} at \SI{322}{\nano\meter} and a beam waist of \SI{1}{\milli\meter}.
    Our sequence, which mimics, for example, a gate sequence in a neutral-atom quantum simulator or computer, establishes piston micro-mirror arrays as a promising platform as a gate controller~\cite{Zhang2024} of atom arrays in the ultra-violet spectral range.
    
    \section{Applications for neutral-atom based quantum technologies}
    In a final set of measurements, we demonstrate the perspectives of using the MMA for quantum simulation or quantum computing.
    In particular, we focus on zone-based architecture recently brought forward~\cite{Bluvstein2022}, and successfully employed for quantum computing~\cite{Bluvstein2023,Radnaev2024,Reichardt2024} and digital quantum simulation~\cite{Evered2025}.
    %
    %
    %
    In this approach, atoms are dynamically shuttled between different zones optimized for storage, manipulation and readout, see~\figref{fig:4}{b}.
    The manipulation zones host multiple atoms to allow for parallel execution of the gate operation, which benefits strongly from a homogeneous illumination of the entire zone.
    This is ideally realized with one or several flat-top beams to minimize the alignment sensitivity of the system and maximize the number of gates executable in parallel.
    Creating such flat-top beams poses a challenge for transitions in the ultraviolet, as found for Rydberg p-states in alkali atoms~\cite{Jau2016a,Zeiher2016b}, and Rydberg s-states in strontium~\cite{Gaul2016,Madjarov2020,Schine2022c} or ytterbium~\cite{Ma2022a,Ma2023,Reichardt2024}, starting from the metastable states.
    We leverage the phase control of the MMA to create flat top beams in the ultraviolet and visible spectral range.
    To compute the phase mask for a flat top beam, we take the measured amplitude on the MMA during the phase-correction measurement as the input field and assume a flat phase profile. 
    We then apply a round aperture on the MMA, directing light outside of the aperture in the zeroth diffraction order of the MMA.
    We then create an amplitude target in the image plane with super-Gaussian decay of order N = 3 in the flanks and a flat amplitude profile at the center.
    With UV laser light, we create a one-dimensional flat-top beam, suitable for executing fast entangling gates across an array of atoms.
    In the visible wavelength range, we also create a two-dimensional flat-top beam with a flat center along both \textit{x}- and \textit{y}-direction with rounded corners which is especially robust against beam pointing fluctuations.
    We optimize the phase profile in a first stage by running a gradient optimization routine minimizing the root mean square-difference between the propagated far-field and the target in a specified region of interest around the flat-top beam. 
    Initially, this procedure results in a flat-top beam with significant intensity variations.
    We therefore image the beam and optimize both the input amplitude for the gradient-descent algorithm, as well as the phase correction by optimizing over the amplitude of Zernike polynomials up to 8-th order.
    By this procedure, we generate flat-top beams at wavelengths of \SI{322}{\nano\meter} and \SI{688}{\nano\meter}, see~\figref{fig:4}. 
    At \SI{322}{\nano\meter} we reach a residual RMS inhomogeneity of $0.75\%$ and peak-to-peak variation of $3.65\%$ over a distance of \SI{113}{\micro\meter} at a power efficiency of $28\%$ which is compatible with reaching state-of-the-art logical gate fidelities across a large number of qubits ~\cite{Jandura2023}. 
    At \SI{688}{\nano\meter} we demonstrate a two-dimensional flattop reaching an RMS inhomogeneity of  $1.8\%$ ($1.4\%$) and peak-to-peak variation of $7.4\%$ ($6.2\%$)  in the \textit{x}- (\textit{y})-directions at a power efficiency of $38\%$.
    \section{Conclusion \& Outlook}
    %
    Our measurements demonstrate that piston micro-mirror arrays can be used for dynamical arbitrary spatial light field control in the visible and UV and are a promising alternative to LCOS-SLMs.
    We reach state of the art performance, both in phase-correction as well as tweezer-homogeneity. 
    Scaling up the number of mirrors would enable even higher diffraction efficiencies for large arrays of thousands of tweezers.
    We show that MMA technology holds promise to enable high-fidelity entangling gates across even larger qubit arrays by generating homogeneous flat-top beams.
    In particular, the current speed limit is not set by fundamental mechanical mirror properties but rather the underlying MEMS-control electronics, which we expect can be updated in a further developed version of the MMA. 
    Beyond the demonstrations shown in this work, the fast update rate of \SI{1}{\kilo\hertz} opens up new possibilities for rapid switching between phase masks with multiple applications for atom-based quantum technologies, such as the recently demonstrated parallel atom resorting~\cite{Kim2016,Lin2024,Knottnerus2025}.

	\begin{acknowledgments}
		%
		We acknowledge funding by the Max Planck Society (MPG) the Deutsche Forschungsgemeinschaft (DFG, German Research Foundation) under Germany's Excellence Strategy--EXC-2111--390814868, from the Munich Quantum Valley initiative as part of the High-Tech Agenda Plus of the Bavarian State Government, and from the BMBF through the programs MUNIQC-Atoms and MAQCS.
		This publication has also received funding under Horizon Europe programme HORIZON-CL4-2022-QUANTUM-02-SGA via the project 101113690 (PASQuanS2.1).
		J.Z. acknowledges support from the BMBF through the program “Quantum technologies - from basic research to market” (SNAQC, Grant No. 13N16265).
		H.T., M.A. and R.T. acknowledge funding from the International Max Planck Research School (IMPRS) for Quantum Science and Technology. M.A acknowledges support through a fellowship from the Hector Fellow Academy.
		F.G. acknowledges funding from the Swiss National Fonds (Fund Nr. P500PT\textunderscore203162).
        O.L. acknowledges support from the Rothschild and CHE Quantum Science and Technology fellowships.
        D.B. and D.K. acknowledge support from the BMBF in the QNC Space project SMAQ as part of the project FMD-QNC (Fund no. 16ME0829K). D.B. and D.K. acknowledge Jörg Heber for his extensive work in SLM development.
        T.R. acknowledges support from the BMBF through the project FMD-QNC.
		
	\end{acknowledgments}
    %
	\appendix

	\bibliography{piston_mems}
	
\end{document}